\documentclass[12pt,draftcls,onecolumn]{IEEEtran}
\usepackage{amssymb}
\usepackage{graphicx}
\usepackage{multicol}
\usepackage{amsmath,epsfig}
\usepackage{amsfonts}
\usepackage{amssymb}
\usepackage{amsthm}
\usepackage{ifthen}
\usepackage{color}
\usepackage{graphicx}
\usepackage{cite}
\usepackage{float}
\usepackage{enumerate}
\usepackage{algorithm}
\usepackage[noend]{algpseudocode}
\makeatletter

\usepackage{epsfig}\usepackage{amsfonts}\usepackage{amsthm}
\usepackage{cite}\usepackage{float}

\makeatother

\begin{document}

\title{Analysis and Throughput Optimization of Selective Chase Combining
for OFDM Systems}

\author{Taniya~Shafique, Muhammad~Zia and Huy-Dung Han\\
 }
\maketitle
\begin{abstract}
In this paper, we present throughput analysis and optimization of bandwidth efficient selective retransmission method 
at modulation layer for conventional Chase Combining (CC) method under orthogonal frequency division multiplexing (OFDM)
signaling. Most of the times, there are fewer errors in a failed packet and receiver can recover from errors receiving 
\emph{partial }copy of original frame. The proposed selective retransmission method at modulation layer for OFDM 
modulation  requests retransmission of information corresponding to the poor quality  subcarriers. In this work, we 
propose cross-layer multiple selective Chase combining (MSCC) method and Chase combining with selective retransmission 
(CCWS) at modulation level.  We also present bit-error rate (BER) and throughput analysis of the proposed MSCC and CCWS 
methods. In order to maximize throughput of the proposed methods under OFDM signaling, we formulate optimization problem  
with respect to amount of information to be retransmitted in selective retransmission in the event of packet failure. We present tight BER upper bounds and  tight throughput lower bounds for the proposed selective Chase combining methods. The simulation results demonstrate significant throughput gain of the optimized 
selective retransmission methods over conventional retransmission methods. The throughput gain of the proposed selective retransmission at modulation layer are also holds for conventional for hybrid automatic repeat request (HARQ) methods.   

\end{abstract}

\section{Introduction}

\label{sec:introduction} 
The contemporary wireless communication
standards such as LTE-advanced \cite{LTEDOc} integrate new technologies
to meet  increasing need of high data rate. The current and future
communication systems employ multiple-input multiple-output (MIMO)
technologies due to their potential to achieve higher data rate and
diversity. In order to assure error-free communication with high throughput,
packet error detection and correction protocols have evolved over
time \cite{ShuLin}. The automatic repeat reQuest (ARQ) method guarantees
error free data transfer using cyclic redundancy check (CRC) approach.
The concept of hybrid ARQ (HARQ) integrates ARQ and forward error correction (FEC)
codes \cite{ShuLin,HARQBroadBand} to provide effective means of enhancing
overall throughput of communication systems. In the event of packet failure, advanced form
of HARQ incorporates joint decoding by combining soft information
from multiple transmissions of a failed packet. Thus, HARQ is one of
the most important technologies embedded in the latest communication
systems such as high-speed down link packet access (HSDPA), universal
mobile telecommunications system (UMTS) that pervade 3G and 4G wireless
networks to ensure data reliability.

In type-I HARQ, receiver request retransmission of a failed packet and discards
observation of the failed packet. Type-II HARQ is most commonly used
method and achieves higher throughput. The type-II HARQ is divided
into Chase combining HARQ (CC-HARQ) and incremental redundancy HARQ
(IR-HARQ). In CC-HARQ, receiver preserves observations of the failed
packet and request retransmission of full packet. The rceiever Chase
combines \cite{ChaseComb} observations of the failed packet and retransmitted
packet to decode packet. In the event of packet failure under IR-HARQ,
receiver request retransmission of more parity information to recover
from error. In response to retransmission request, transmitter sends
more parity bits lowering code rate of FEC. The receiver combines
new parity bits with buffer observation for FEC decoding.

Most of the research conducted on HARQ focused on ARQ and FEC \cite{HARQBroadBand,ShenFitz}
without exploring modulation layer. Throughput of capacity achieving codes FECs such as LDPC and 
Turbo codes is optimized for Rayleigh fading channel  in \cite{ARQJindal} with ARQ and HARQ protocols. 
Mutual information based performance analysis for HARQ over Rayleigh fading channel is provided in 
\cite{JindalHARQAnalyRayFad}.  Optimal power allocation for Chase combining based HARQ is 
optimized in \cite{OptPowerLarson,OutageLarson, AdapPowerLarson,EngAwareShami,GreenHARQ}. Without exploiting channel state
information and frequency diversity of the frequency selective channel,
partial retransmission of the original symbol stream of a failed packet
is addressed in \cite{ZiaDingGLOBECOM2008,ZhiDingPiggy,ZiaDingHARQOSTBC}.
These methods retransmit punctured packet in predetermined fashion.
Furthermore, the complexity of joint detection for \emph{partial }retransmission
is \cite{ZhiDingPiggy,ZiaDingGLOBECOM2008,Jermey2007} not tractable.
Partial retransmission of orthogonal space-time block (OSTB) coded
\cite{Alamouti} OFDM signaling is proposed in \cite{ZiaDingHARQOSTBC}.
In \cite{ARQMinitor}, for conventional ARQ protocol, full packet
retransmission at modulation layer is employed when channel gain is
below a threshold value without buffering observations of low SNR channel
realization. In a typical failed packet, there are small number of
corrupted bits and retransmission of full packet is not necessary.
The receiver can recover from errors by retransmission of potentially
culprit bits. The OFDM signaling allows to identify poor quality bits
corresponding to the subcarriers that have low signal-to-noise ratio
(SNR). Selective retransmission at modulation layer of OFDM signaling
proposed in \cite{ZiaSelHARQ,ETRIMIMOCondNum,IETOFDMPartial,SelecCCforOSTBC} 
achieves throughput gain as compared
to conventional HARQ methods. Throughput optimization of selective
retransmission and performance analysis is not addressed in \cite{ZiaSelHARQ}.

The motivation of selective retransmission owing to the fact that in the event of failed 
packet under OFDM signaling at MAC layer, often receiver can recover from error by retransmitting 
\emph{partial} information  corresponding to the poor quality subcarriers. An OFDM signaling allows 
 retransmission of information transmitted over poor quality subcarrier at physical layer (PHY)  selectively. 
After receiving copy of information symbols corresponding to poor quality subcarriers, receiver jointly 
decode data in Chase combining fashion. In this work, we propose low complexity and bandwidth efficient 
multiple selective Chase combining (MSCC) and Chase combining with selective retransmission (CCWS) 
methods in cross-layer fashion at modulation layer for OFDM signaling. We also 
provide  bit-error rate (BER)  and throughput analysis in terms tight upper BER bound and lower throughput 
bound, respectively for the proposed retransmission schemes. The amount of information to be retransmitted for 
each subcarrier in the event of failed packet is function of signal-to-noise ratio (SNR) of the corresponding 
subcarrier. Using norm of  channel gain for each subcarriers as channel quality measure, we also optimized 
threshold $\tau$ on channel norm for selective retransmission in order to maximize throughput. The simulation 
results demonstrate that proposed methods offers substantial throughput gain over conventional CC method 
in low SNR regime. Our results also show that there is marginal gap between analytical bounds and simulation 
results (Monte Carlo method ) for BER and throughput of the proposed methods. The throughput 
gain of the proposed schemes also hold with FEC.

We organize our manuscript as follows. First, we present system model and problem formulation (MSCC and CCWS)
for selective Chase combining of OFDM system in Section~\ref{sec:SystemModel} and Section~\ref{subProblemFormulation}.
In Section~\ref{sec:perfomnceanlysis}, we present BER analysis of
MSCC for one selective retransmission at modulation layer and CCWS methods.  
Throughput analysis for SCC and CCWS are presented in
Section~\ref{sec:throughAnalysis}. Throughput Optimization is performed
in Section~\ref{sec:OptThrough}. We discuss our results in Section~\ref{Sec:simulation}.
Finally, we conclude the purposed work in Section~\ref{Sec:CONCLUSION}.

\section{SYSTEM MODEL and Problem Formulation}
\subsection{System Model}
\label{sec:SystemModel}
We consider OFDM system for selective
retransmission with $n_{r}$ receive antennas. Transmitter transmits
information symbol using OFDM signaling over frequency selective channel
of $L$ coefficients. An OFDM signaling converts frequency selective channel 
$\mathbf{h}_{i}$, where $i=1,\cdots,n_{r}$, from transmit antenna
to the receive antenna $i$ into $N_{s}$ parallel channel\cite{Larsson}.
Thus, a channel gain vector of $\ell$-th subcarrier between transmitter
and receiver pair is $H(\ell)=\big[H_{1}^{T}(\ell)\,\,\cdots H_{n_{r}}^{T}(\ell)\big]^{T}$,
where elements of channel vector $H(\ell)$ are generated by applying
Fourier transformation matrix $F\in\mathcal{C}^{N_{s}\times N_{s}}$
on frequency selective channel $\mathbf{h}_{i}$. The elements of
vector $H(\ell)$ are independent and identically distributed (i.i.d.) with distribution $\mathcal{N}_{c}(0,1)$
\cite{Larsson}. Note that for $n_{r}=1,$ SIMO-OFDM system becomes
SISO-OFDM system. The matrix model of the received vector $\mathbf{y}(\ell)$
for the $\ell$-th subcarrier can be written as 
\begin{align}
\mathbf{y}(\ell)=H(\ell)\cdot s(\ell)+\mathbf{w}(\ell),
\end{align}
where vector $\mathbf{w}(\ell)\sim\mathcal{N}_{c}(0,N_{0}I)$ is an additive white
Gaussian noise vector. 
A typical failed packet has few erroneous bits
and bits transmitted over OFDM subcarriers with small channel norm
$\Vert H(\ell)\Vert^{2}$ are more susceptible to the impairments.
OFDM signaling allows retransmission of targeted information symbols corresponding to the
poor quality subcarriers instead of unnecessary retransmission of
full packet \cite{ZiaSelHARQ}. The joint detection by combining observation
of the first transmission and subsequent selective retransmission
of the poor quality subcarriers enhances throughput of the transceiver
under OFDM signaling. The selective retransmitted information improves
bit-error rate of the poor quality subcarriers resulting into throughput
gain. Note that retransmission of more information does not increase
throughput linearly. The amount of information to be retransmitted
is controlled by threshold $\tau$ on the channel norm $\Vert H(\ell)\Vert^{2}$
of the $\ell$-th subcarriers. The optimization of threshold $\tau$
in order to maximize throughput $\eta$  of the transceiver under selective retransmission is the main focus of this
work. Throughput of selective retransmission is function of probability
of error, which in fact is function of $\tau$. We present uncoded BER analysis
and throughput analysis in Section \ref{sec:perfomnceanlysis} and
\ref{sec:throughAnalysis}, respectively, in order to optimize parameter
$\tau$ in Section \ref{sec:OptThrough}.  Next, we provides description of proposed MSCC and 
CCWS methods.
\begin{figure}[t]
\centering \includegraphics[width=12.3cm]{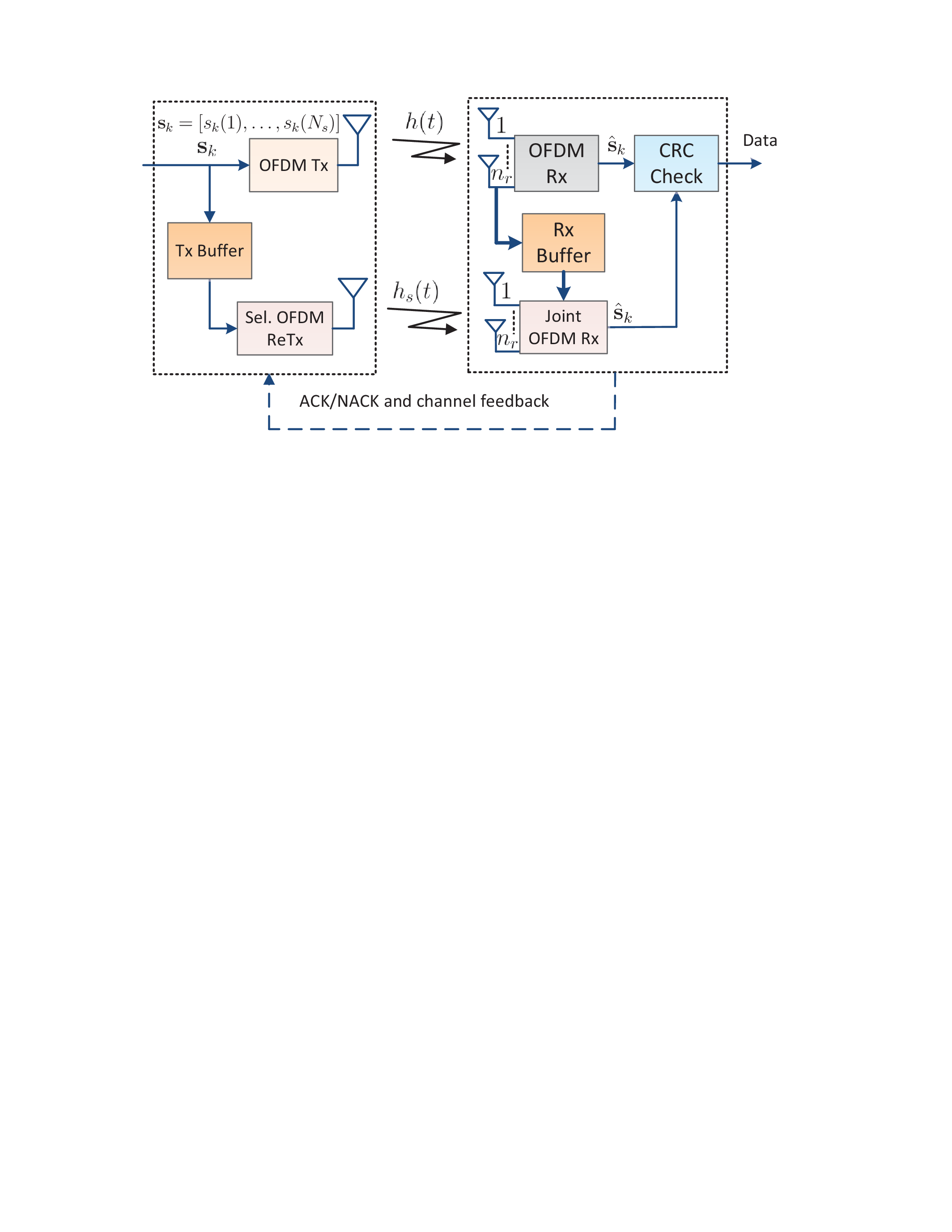} 
\caption{ Selective retransmission model for SIMO-OFDM system at modulation layer} 
\label{fig:SystemModelSCC}
\end{figure}
\subsection{Problem Formulation}
\label{subProblemFormulation}
Now we present  our proposed cross-layer MSCC and  CCWS methods  for OFDM 
signaling. In both MSCC and CCWS methods, medium access control (MAC) layer initiates 
retransmission in the event of CRC failure. Based on quality of the subcarriers, instead of full retransmission under convetional Chase combining, PHY layer initiates selective 
retransmission of information symbols corresponding to the poor quality subcarriers along with NACK signal. 
We use norm of subcarrier gain vector $\Vert {H(\ell)}\Vert^2$ as channel quality measure.
Note that OFDM signaling allows retransmission of information symbols transmitted over poor 
quality subcarriers ($\Vert {H(\ell)}\Vert^2< \tau$ ) selectively avoiding overhead of retransmission 
of information symbols corresponding to good quality subcarriers, where $\tau$ is threshold on 
channel norm.  The receiver feeds back partial 
channel state information (PCSI) when  each coherent time is elapsed. We assume that due to 
longer retransmission delay, each retransmission encounters independent channel. First, we present MSCC method under OFDM signaling.
\subsubsection {Multiple selective Chase combining}
\label{subsecMSCC}

Motivation of selective Chase combining (SCC) stems from the fact that under OFDM signaling, in the event of CRC failure,  receiver can recover from errors by selectively retransmitting information symbols corresponding to the poor quality subcarriers without changing modulation. In  SCC, instead of requesting full retransmission for conventional Chase combining, PHY layer requests retransmission of information symbols corresponding to the poor quality subcarriers and retains observation of the failed data in a buffer. In response to NACK signal, transmitter retransmits requested information symbols. The receiver combines observation of the first transmission and selective retransmission of the poor quality subcarriers for joint decoding  in Chase combining fashion and discards buffered observations.This is defined as one round of SCC. Note that one round of conventional CC consists of first transmission followed by full retransmission in the event of CRC failure. The receiver decodes information from observations of first transmission and selective retransmission  jointly. Let $\mathcal{M}$ be the maximum allowed number of rounds of CC at MAC layer.  At the end of each retransmission round of CC, receiver discards buffered observations and initiates new CC round (if CRC check fails). In selective retransmission, the amount of information to be retransmitted is controlled by threshold $\tau$ on the norm of subcarrier gain $\Vert {H(\ell)}\Vert^2$. The threshold $\tau$ can be optimized to maximized throughput for given average SNR. In SCC, one round of selective retransmission consists of one round of CC. In \cite{ZiaSelHARQ} and \cite{SelecCCforOSTBC}, SCC method is presented for SISO-OFDM and MIMO-OFDM under OSTB coded signals, respectively. In this work, we optimize threshold $\tau$ on channel quality to maximize throughput for SCC and present throughput analysis. We also propose MSCC method that achieves significant throughput gain over SCC method. We present provide tight upper BER performance bound and throughput analysis of proposed MSCC method. 
 
One retransmission round of MSCC method in PHY consists of  at most $\Omega$ many conventional CC rounds at MAC layer. In one round of  MSCC method, a receiver continues to request retransmission of information symbols over $\ell$th subcarrier of a failed packet until sum of norm-square of the subcarrier gains from all iterations of selective retransmission of MSCC method is larger than threshold $\tau$ or allowed number of retransmissions is reached. If CRC check is satisfied, selective retransmission and CC rounds are terminated followed by ACK signal to the transmitter. Let $J$ be the count of retransmission rounds of CC at MAC layer.  For the $I$-th iteration of selective retransmission at modulation layer of the $J$-th retransmission round at MAC layer, if needed at modulation layer, receiver request retransmission of poor quality subcarrier which have $\sum_{i=1}^I \Vert H_i(\ell) \Vert^2< \tau$ for $I=1,\hdots,\Omega$, where $\Omega$ is the maximum number of retransmissions considered for joint detection in $J$-th transmission round at MAC layer. At the end of each conventional CC round at MAC layer, when MSCC method is enabled, receiver jointly detects information bits by combining observations of all previous $I$ selective retransmissions iterations. After $I=\Omega$ iterations of selective retransmission, which constitute one round of proposed MSCC method, receiver discards buffered observation and initiates new selective retransmission round. 
  
Note that $I=1$ represents first transmission of a packet and $\mathcal{T}_{I}$ is set of indices of poor quality subcarriers of the $I$-th selective retransmission. The set of indices of poor quality subcarriers of $I+1$-th selective retransmission $\mathcal{T}_{I+1}$ is subset of  set of poor quality indices $\mathcal{T}_{I}$ of the $I$-th selective retransmission. That is, $\mathcal{T}_{I+1} \subseteq \mathcal{T}_I$.   The receiver stops retransmission of information symbols corresponding to the subcarriers in the iterations of selective retransmission at modulation layer for which $\sum_{i=1}^I \Vert H_i(\ell) \Vert^2\ge \tau$.  Let $\beta_I$ be the cardinality $\vert \mathcal{T}_{I}\vert $ of set $\mathcal{T}_{I}$, which represents number of poor subcarriers for the $I$-th iteration of selective retransmission of the $J$-th round at MAC layer. The flow graph of MSCC  selective retransmissions method is presented in Algorithm~\ref{AlgoMSCCRevised}.
\begin{algorithm}
\caption{MSCC protocol}
\label{AlgoMSCCRevised}
\begin{algorithmic}[1]
\State   $J=1$ corresponds to the first transmission of the $k$-th packet
\State Set $I=1$
\State  Detection and CRC check of first iteration of MSCC at PHY layer and $J$-th round at MAC layer of $k$-th packet 
          \If {CRC satisfies} $k=k+1$, discard observations and go to 1
					\EndIf
  \If {$I= \Omega$} go to 9
	\EndIf
\State    Selective retransmission for the $\beta_I$ subcarriers which have $\displaystyle \sum_{i=1}^I \Vert H_i(\ell) \Vert^2< \tau$ and  preservation of  observations 
\State Joint detection from $\displaystyle N_s+\sum_{i=1}^I \beta_i$ observations
\State Set $I=I+1$, $J=J+1$ and go to 4
  \If {$J= {\mathcal{M}}$}, discard observations, declare packet loss, $k=k+1$ and go to 1
	\EndIf 
\State   Discard observations and go to 2
\end{algorithmic}
\end{algorithm}
We present tight upper bound on uncoded BER and lower bound on throughput of MSCC method under OFDM signaling over Rayleigh fading channel in Section~\ref{sec:perfomnceanlysis} and Section~\ref{subsec:throughAnalsisSCC}, respectively. We also optimized threshold on channel quality that controls amount of information need to be transmitted that maximizes throughput in Section~\ref{sec:OptThrough}.  
\subsubsection {Chase combining with selective retransmission}
\label{subsecCCWS}

Now we discuss proposed CCWS protocol for OFDM signaling. In proposed CCWS method, MAC layer, similar to conventional CC method,   initiates full retransmission in the event of packet failure. The proposed CCWS method is different from convention CC method in the sense that CCWS method employs a selective retransmission corresponding to the  poor quality subcarriers at PHY level  independent of MAC layer for the first transmission and full retransmission. That is, first transmission of each packet and full retransmission of a failed  packet both undergo a selective retransmission of the information symbols corresponding to the poor quality subcarriers at PHY layer.  The channel quality ($\Vert H(\ell)\Vert^2$) of subcarriers for the first transmission of a packet under CCWS method is compared with threshold $\tau$ prior to decoding at PHY layer. The receiver request selective retransmission request to the transmitter and preserves observations of the first transmission of a packet. Then, receiver decodes information by fusing observations of the first transmission and selective retransmission jointly. If CRC check is not satisfied, receiver buffers observation and requests full retransmission of failed packet by sending NACK signal. Similar to the first transmission, receiver initiates selective retransmission of the information symbols corresponding to poor quality subcarriers of full retransmission for which $\Vert H_c(\ell)\Vert^2 < \tau$. Note that $H(\ell)$ and  $H_c(\ell)$ are the gains of $\ell$-th subcarrier for the first transmission and full retransmission, respectively. One round of CCWS consists of a full transmission, a full retransmission and two selective retransmissions.  After response of the NACK signal from transmitter, the receiver decodes information by combining observation of the one round of CCWS. If CRC check is satisfied, receiver send ACK signal to the transmitter. Otherwise, receiver clears buffered observations and initiates next round of CCWS. When maximum number of rounds of CCWS are elapsed, packet failure is declared to RLC layer. In order to maximize throughput of CCWS, we optimizes threshold $\tau$ on channel quality. 

Let $\beta_1$ be the number of  subcarriers with channel gains $\Vert H(\ell)\Vert^{2} <\tau$  for the first transmission of CCWS method. The modulation layer initiates selective retransmission of symbols corresponding to the $\beta_1$ subcarriers prior to joint detection. Due to selective retransmission and joint detection at modulation layer, BER performance $P_{e_1}$ of the first transmission of CCWS method at  MAC layer is same as that of MSCC method with $\Omega=1$.

 Note that channel vector of the poor quality subcarriers of OFDM signaling of CCWS method of
the first transmission as a result of selective retransmission is stack of two channel realizations similar
to SCC. The decoded  $N_s\mbox{log}_2 M$ bits are processed for  CRC check.  In the event of  CRC failure for the first transmission of a packet, receivers preserves observations of the first transmission ($\beta_1+N_s$ symbols)
 and initiates full retransmission request of packet.  The modulation layer initiates selective retransmission of  $\beta_2$ poor quality subcarrier for which $\Vert H_c(\ell)\Vert^2 < \tau$ similar to the first transmission, where $E\left[\beta_1\right ]=E\left[\beta_2\right] =P(\Vert H_1(\ell)\Vert^2 < \tau)=P(\Vert H_c(\ell)\Vert^2 < \tau)=m$.  The joint detection as a result of full retransmission at MAC layer processes $\beta_1+\beta_2+2N_s$ observations to decode $N_s\mbox{log}_2 M$ bits  and achieves BER performance $P_{e_2}$. Let $\mathcal{M}$ be the maximum number of allowed rounds of transmissions of a packet and $J$ be the  round counter for the transmission of the $k$-th packet of  CCWS method. The flow graph of CCWS  protocol for the $k$-th packet is presented in Algorithm~\ref{AlgoCCWSRev}. We present tight upper bound on uncoded BER and lower bound on throughput of CCWS method under OFDM signaling over Rayleigh fading channel in Section~\ref{SubBERCCWS} and Section~\ref{subsec:throughAnalsisCCWS}, respectively. We also optimized threshold on channel quality that controls amount of information need to be transmitted that maximizes throughput in Section~\ref{sec:OptThrough}.  
\begin{algorithm}
\caption{CCWS protocol}
\label{AlgoCCWSRev}
\begin{algorithmic}[1]
\State   $J=1$ corresponds to the first transmission of the $k$-th packet
\State  Selective retransmission of $\beta_1$ subcarriers and preservation of $N_s$ observations
\State   Joint decoding from $N_s+\beta_1$ observations and  CRC check
 \If {CRC  satisfies}  $k=k+1$, discard observations and go to 1
            \EndIf
\State   NACK signal for full retransmission of packet at MAC layer and preservation $\beta_1+N_s$ observations.
\State   Selective retransmission of $\beta_2$ poor quality subcarriers corresponding to full retransmission and preservation of $N_s$ observations
\State Joint detection from  $\beta_1+\beta_2+2N_s$ observations
          \If {CRC satisfies} $k=k+1$, discard observations and go to 1
					\EndIf
   \If {$J> \mathcal{M}$} declare packet  loss, discard observation, $k=k+1$ and go to 1 \EndIf
\State   $J=J+1$, discard observation and go to 2
\end{algorithmic}
\end{algorithm}

Performance analysis of SCC and CCWS is presented next. 
\section{Performance Analysis}
\label{sec:perfomnceanlysis} In this section, we present tight upper bounds on BER 
of joint detection for MSCC and CCWS methods at PHY layer. We
use these tight upper bound on BER to optimize throughput of the proposed selective retransmission
methods in Section \ref{sec:throughAnalysis}. We first derive upper bound on BER of MSCC method discussed in Section~\ref{subsecMSCC} for $\Omega=1$ (SCC). 

\subsection{BER analysis of SCC}
\label{subsec:BERAnaSCC}
One round of MSCC consists of $\Omega$ round of conventional CC  method. In case of CRC failure of the first transmission,
 receiver buffers observations
corresponding to the first transmission and requests retransmission
of the poor quality subcarrier as shown in Figure~\ref{fig:SystemModelSCC}. 
Let $h(t)$ and $h_s(t)$ be the channel impulse responses of the first transmission 
and selective retransmission, respectively shown in 
Figure~\ref{fig:SystemModelSCC}. We 
assume that $h(t)$ and $h_s(t)$ are independent due to longer retransmission delay. Note that norm of subcarrier gain and
norm of subcarriers channel vector are channel quality measures for
SISO-OFDM and SIMO-OFDM systems, respectively. In a typical failed
packet, there are fewer error bits and retransmission of potentially
culprit bits corresponding to the poor quality sub-carries can help
receiver to recover from error by performing joint detection. In SCC,
PHY layer is aware of retransmission due to cross-layer design and performs joint detection
for poor quality subcarriers combining observations from the first
transmission and subsequent selective retransmission.

Each element of the complex channel vector $H(\ell)$ of the $\ell$-th subcarrier
follows Gaussian distribution with mean zero and unit variance obtained by applying $N_s$ point DFT matrix on frequency selective channel $h(t)$ of the first transmission of a CC round .
The estimate of information symbol over the $\ell$-th subcarrier
after equalization is 
\begin{align}
\hat{s}(\ell)=s(\ell)+\frac{H^{*}(\ell)\mathbf{w}(\ell)}{\|H(\ell)\|^{2}}=s(\ell)+\mathbf{u}(\ell),
\label{eq:ZFSingleTx}
\end{align}
where $\mathbf{u}(\ell)$ is the effective noise with distribution
$\mathcal{N}(0,\|H(\ell)\|^{-2}N_{0})$. Note that channel $H(\ell)$
and additive noise $\mathbf{w}(\ell)$ are scalars and vectors for SISO-OFDM
and SIMO-OFDM systems, respectively. Selective retransmission method
at modulation layer retransmits \emph{partial copy} of symbols by
targeting symbols corresponding to the poor quality subcarriers for
retransmission. Thus, information symbols corresponding to the subcarriers
with channel norm-square $\|H(\ell)\|^{2}$ less than threshold $\tau$ are selected for retransmission. That is, $ (\|H(\ell)\|^{2}\leq\tau)$.  The symbols transmitted over
subcarriers for which $\|H(\ell)\|^{2}>\tau$ are omitted from retransmission
due to the fact that they are less susceptible to impairments. In
this work, we search for threshold $\tau$, which optimizes the amount
of information to be retransmitted in order to maximize throughput
of an OFDM system.

We denote the outcomes $\|H(\ell)\|^{2}>\tau$ and $\|H(\ell)\|^{2}\le\tau$
by the events $\xi$ and $\xi^{c}$, respectively. The probabilities of events $\xi$ and $\xi^{c}$ are
\begin{align*}
P(\xi^{c}) & =P(\|H(\ell)\|^{2}\le\tau)\;\mbox{and}\; P(\xi)=P(\|H(\ell)\|^{2}>\tau),
\end{align*}
respectively, where random variable $\chi_{1}=\Vert H(\ell)\Vert^{2}$ has Chi-square
distribution of degree $2n_{r}$ \cite{DgitalCommProakis} and $P(\xi^{c})=1-P(\xi)=P(\chi_{1}\le\tau)$
. Now for Rayleigh fading channel, 
\begin{align}
P(\xi^{c}) & =1-\exp\left(-\frac{\tau}{2\sigma^{2}}\right)\sum_{k=0}^{n_{r}-1}\frac{1}{k!}\,\left(\frac{\tau}{2\sigma^{2}}\right)^{k},\\
P(\xi) & =P(\chi_{1}>\tau)=\exp\left(-\frac{\tau}{2\sigma^{2}}\right)\sum_{k=0}^{n_{r}-1}\frac{1}{k!}\,\left(\frac{\tau}{2\sigma^{2}}\right)^{k},
\end{align}
where $\sigma^{2}=\frac{1}{2}$ is variance of real and imaginary components of complex channel coefficient of a subcarrier.
When event $\xi^{c}$ occurs, receiver performs joint detection by
combing observation of the first transmission and subsequent retransmission.
The combined channel response $\mathcal{H}(\ell)=\big[H^{T}(\ell)\,\, H_{s}^{T}(\ell)\big]^{T}$
is constructed by stacking channel of the first transmission and selective
retransmission. If there are $\beta_1$  many subcarriers with $\Vert H(\ell)\Vert^2\le \tau$, then there will be joint detection for $\beta_1$ subcarrier for SCC method ($\beta_1=N_s\implies \mbox{full retransmission}$). Estimate of information symbol as a result of joint
detection is 
\begin{align}
\hat{s}(\ell)= & s(\ell)+\underbrace{\Vert\mathcal{H}(\ell)\Vert^{-2}\mathcal{H}^{H}(\ell)\tilde{\mathbf{w}}(\ell)}_{\tilde{\mathbf{u}}(\ell)},\label{eq:SelDetector}
\end{align}
where $\tilde{\mathbf{u}}(\ell)\thicksim\mathcal{N}_{C}(0,\Vert\mathcal{H}(\ell)\Vert^{-2}N_{0})$
and $\tilde{\mathbf{w}}(\ell)\thicksim\mathcal{N}_{C}(0,\, N_{0}I_{n_{r}})$.
Note that the random variable $\Vert\mathcal{H}(\ell)\Vert^{2}$
also has Chi-square distribution of degree $4n_{r}$. Also that $\Vert\mathcal{H}(\ell)\Vert^{2}=\chi_{1}+\chi_{2}$, 
where Chi-square random variables $\chi_{1}$ and $\chi_{2}=\Vert H_{s}(\ell)\Vert^{2}$
are i.i.d. of degree $2n_{r}$ each. The bit-error probability
of joint detection for selective retransmission over Rayleigh fading channel is
\begin{align}
P_{e_{s}} & =E_{H}\Big [P(\xi^{c})\,\,\, P_{e\vert\xi^{c}}+P(\xi)\,\,\, P_{e\vert\xi}\Big ],
\label{eq:AvgOverH}
\end{align}
where $P_{e\vert\xi}$ and $P_{e\vert\xi^{c}}$ are the conditional bit-error probabilities of single detection and joint detection,
respectively. Now we evaluate threshold $\tau$ to achieve target BER. Let $P_{e\tau}$  be the bit-error rate  and $\tau$ be the corresponding channel norm of a subcarrier. Then, for constellation set of cardinality $M$, relationship between $P_{e\tau}$ and $\tau$ is  
\begin{align}
P_{e\tau}= c\cdot Q\left(\sqrt{g\frac{\tau}{N_{0}}}\right)\Longrightarrow \tau=\frac{N_{0}}{g}\left(Q^{-1}\left(\frac{P_{e\tau}}{c}\right)\right)^{2},
\end{align}
where $c$ and $g$ are modulation constants of constellation set \cite{Larsson}. Since $Q(.)$ is monotonically 
decreasing function SNR, in order to achieve BER smaller than $P_{b\tau}$, $\Vert H(\ell)\Vert^2 > \tau$. That is, $P_{e\vert \xi} \le P_{e\tau}$. Therefore, threshold on the gain  of a subcarrier is
\begin{align}
\Vert H(\ell)\Vert^{2} & \ge \tau=\frac{N_{0}}{g}\left(Q^{-1}\left(\frac{ P_{b\tau}}{c}\right)\right)^{2}.
\label{eq:ThresholdTau}
\end{align}
We use this threshold for selective retransmission at modulation layer to ensure $P_e \le P_{e\tau}$. In Section~\ref{sec:OptThrough}, we search for threshold $\tau$  that maximized throughput . 
The upper bound on probability of error for joint detection of selective Chase combining
(SCC) is 
\begin{align}
P_{e_{s}} & = P(\xi^{c})\cdot c\cdot E_{H\vert\xi^{c}}\left[Q\left(\sqrt{g\frac{\chi_{1}}{N_{0}}}\right)\right]+ P(\xi)\cdot c\cdot E_{\mathcal{H}\vert\xi}\left[Q\left(\sqrt{g\frac{\chi_{1}+\chi_{2}}{N_{0}}}\right)\right],
\label{eq:ProbOfError}
\end{align}
where $E_{H\vert\xi^{c}}$ and $E_{\mathcal{H}\vert\xi}$ are conditional
expectations. The subscript $s$ in  \eqref{eq:ProbOfError} represent selective Chase combining method. The conditional probability density function $f_{\chi_1\vert\xi^c}(x_1)$  of $f_{\chi_1}(x_1)$ when $\chi_1>\tau$ is $f_{\chi_1\vert\xi^c}(x_1) =\dfrac{f_{\chi_1}(x_1)}{P(\xi^c)}, \mbox{ where } \chi_1 >\tau$. In order to solve first term of \eqref{eq:ProbOfError}  \cite{StarkWoods } , we have
\begin{align}
\begin{split}
\displaystyle  E_{H\vert\xi^{c}}\left[Q\left(\sqrt{g\frac{\chi_{1}}{N_{0}}}\right)\right]=&\int_{\tau}^{\infty} Q \left(\sqrt{g\frac{x_{1}}{N_{0}}}\right) \dfrac{f_{\chi_1}(x_1)}{P(\xi^{c})} dx_1.
\end{split}
\label{eq:SCC2}
\end{align}
Similarly,
\begin{align}
\begin{split}
\displaystyle  E_{H\vert\xi}\left[Q\left(\sqrt{g\frac{\chi_{1}+\chi_2}{N_{0}}}\right)\right]=
&\int_{x_1=0}^{\tau} \int_{x_2=0}^{\infty}Q\left(\sqrt{g\frac{x_{1}+x_2}{N_{0}}}\right)  \dfrac{f_{\chi_1}(x_1)f_{\chi_2}(x_2)}{P(\xi)} dx_2 dx_1.
\end{split}
\label{eq:SCC3}
\end{align}
The upper bound on $P_{es}$ in \eqref{eq:ProbOfError} using approximation
of Q-function \cite{QFuncApprox}, \eqref{eq:SCC2} and \eqref{eq:SCC3} can be written as \cite{SelecCCforOSTBC,StarkWoods } 
\begin{align}
\begin{split}
P_{e_s}\le&\frac{c}{12} \int_\tau^\infty \exp(-g\frac{x_1}{2N_0})  f_{\chi_1}(x_1) dx_1	
+\frac{c}{4} \int_\tau^\infty \exp(-g\frac{4 x_1}{3.2N_0})  f_{\chi_1}(x_1) dx_1\\	
&+\frac{c}{12} 	\int_0^\tau \exp(-g\frac{x_1}{2N_0})  f_{\chi_1}(x_1) dx_1	.\int_0^\infty \exp(-g\frac{x_2}{2N_0})  f_{\chi_2}(x_2) dx_2\\
	&+\frac{c}{4} 	\int_0^\tau \exp(-g\frac{4x_1}{3.2N_0})  f_{\chi_1}(x_1) dx_1	.\int_0^\infty \exp(-g\frac{4x_2}{3.2N_0})  f_{\chi_2}(x_2) dx_2
	\label{eq:UBSCC}
\end{split}
\end{align}
 Simplifying \eqref{eq:UBSCC}, we have upper bound on probability of error of the joint detection as follows
\begin{align}
\begin{split}
P_{e_{s}}  \le&\frac{c}{12}\left(\frac{\rho}{\sigma}\right)^{2n_{r}}\exp\left(-\frac{\tau}{2\rho^{2}}\right)\sum_{k=0}^{n_{r}-1}\frac{1}{k!}\left(\frac{\tau}{2\rho^{2}}\right)^{k}+\frac{c}{4}\left(\frac{\rho_{1}}{\sigma}\right)^{2n_{r}}\exp\left(-\frac{\tau}{2\rho_{1}^{2}}\right)\sum_{k=0}^{n_{r}-1}\frac{1}{k!}\left(\frac{\tau}{2\rho_{1}^{2}}\right)^{k}\\
 & +\frac{c}{12}\left(\frac{\rho}{\sigma}\right)^{4n_{r}}\left(1-\exp\left(-\frac{\tau}{2\rho^{2}}\right)\sum_{k=0}^{n_{r}-1}\frac{1}{k!}\left(\frac{\tau}{2\rho^{2}}\right)^{k}\right)\\
&+\frac{c}{4}\left(\frac{\rho_{1}}{\sigma}\right)^{4n_{r}}\left(1-\exp\left(-\frac{\tau}{2\rho_{1}^{2}}\right)\sum_{k=0}^{n_{r}-1}\frac{1}{k!}\left(\frac{\tau}{2\rho_{1}^{2}}\right)^{k}\right),
\end{split}
\label{eq:BERUpperBoundJointSCC}
\end{align}
\noindent
where $\rho=\sqrt{\frac{\sigma^2 N_{o}}{g\sigma^2+N_{o}}}$, $\rho_{1}=\sqrt{\frac{\sigma^{2}N_{o}}{g_{1}\sigma^{2}+N_{o}}}$
and $g_{1}=\frac{4g}{3}$. For 4-QAM constellation $g=2$ and $c=\frac{2}{\text{log}_2M}$
\cite{DgitalCommProakis}. Note that in \eqref{eq:BERUpperBoundJointSCC},
$P_{e_j}$ is function of $\tau$, which controls amount of information
to be retransmitted in selective retransmission.The probability
of error of the first transmission with $n_{r}$ receiver antenna can
be derived from \eqref{eq:BERUpperBoundJointSCC} by setting parameter
$\tau\rightarrow 0$, which means that probability of selecting
a subcarrier for retransmission is almost zero. For $\tau\rightarrow 0$,
\eqref{eq:BERUpperBoundJointSCC} becomes
\begin{align}
\begin{split}P_{e}\le & \frac{c}{12}\left(\frac{\rho}{\sigma}\right)^{2n_{r}}+\frac{c}{4}\left(\frac{\rho_{1}}{\sigma}\right)^{2n_{r}},
\end{split}
\label{eq:BERUpperBoundSingleSCC}
\end{align}
which is probability of error of the first transmission with $n_r$ receive antennas. Similarly, for full retransmission, $\tau\rightarrow \infty$ and BER upper bound can be achieved by setting $\tau=\infty$ in \eqref{eq:BERUpperBoundJointSCC}.  Thus,
\begin{align}
\begin{split}P_{e_f}\le & \frac{c}{12}\left(\frac{\rho}{\sigma}\right)^{4n_{r}}+\frac{c}{4}\left(\frac{\rho_{1}}{\sigma}\right)^{4n_{r}}.
\end{split}
\label{eq:BERUpperBoundFullCC}
\end{align}
In low SNR regime, receiver needs more information during retransmission
to recover from errors. Therefore, threshold $\tau$ on channel norm
has large value in low SNR regime. Throughput $\eta$ is function
of probability of error of first transmission $P_e$ and probability of error of joint detection $P_{e_s}$, which is function of SNR and threshold $\tau$. We use \eqref{eq:BERUpperBoundJointSCC} and
\eqref{eq:BERUpperBoundSingleSCC}  
 to search of optimal value
of $\tau$ to  maximize throughput of SCC method in Section~\ref{sec:throughAnalysis} .
Now we provide BER analysis of CCWS method.

\subsection{BER analysis of CCWS}
\label{SubBERCCWS}
The probability of bit-error $P_{e_1}$ of the first transmission of CCWS is
in fact $P_{e{}_{s}}$ for SCC given in \eqref{eq:BERUpperBoundJointSCC}.
There are two outcomes from the first transmission related to the
$\ell$-th OFDM subcarrier denoted by $\xi$ and $\xi^{c}$ with probabilities
$P(\xi)=P(\Vert H(\ell)\Vert^{2}>\tau)$ and $P(\xi^{c})=P(\Vert H(\ell)\Vert^{2}\le\tau)=1-P(\xi)$
similar to SCC. It is important to note that BER $P_{e_1}$of the first transmission
of CCWS is upper bounded by \eqref{eq:BERUpperBoundJointSCC}. That
is, $P_{e_1}=P_{e_{s}}$ .

Now we evaluate upper bound on BER for joint detection when CRC failure
occurs for CCWS method. Let $H_{c}(\ell)$ be the channel of the $\ell$-th
subcarrier during full retransmission of packet of Chase combining
protocol. Similar to SCC method, modulation layer initiates retransmission 
of the poor subcarriers $\Vert H_{c}(\ell)\Vert^{2}<\tau$ for the full retransmission. We denote the event $\Vert H_{c}(\ell)\Vert^{2}\le\tau$
by $\xi_{c}^{c}$. The subscript $c$ represents retransmission of
Chase combining and superscript with event $\xi_{c}$ represent complement
of $\xi_{c}$. The channel realizations during first transmission
and conventional Chase combining are $H(\ell)$ and $H_{c}(\ell)$, respectively,
are i.i.d. and $P(\xi)=P(\Vert H(\ell)\Vert^{2}>\tau)=P(\xi_{c})=P(\Vert H_{c}(\ell)\Vert^{2}>\tau)$.
In response to full retransmission of failed packet, receiver combines
$\beta_s+N_s$ observations of the first transmission and $\beta_c+N_s$ of full retransmission for joint
detection. Note that $\beta_c$ is the number of poor subcarriers retransmitted at modulation layer due to full retransmission at data link layer. Probability of error of joint detection is as follows:
\begin{align}
\begin{split}
P_{e_{2}}  =&E_{H}\big[P(\xi_{1})P_{e\vert\xi_{1}}+P(\xi_{2})P_{e\vert\xi_{2}}+
 P(\xi_{3})P_{e\vert\xi_{3}}+P(\xi_{4})P_{e\vert\xi_{4}}\big],
\label{eq:CCWSjoitBER}
\end{split}
\end{align}
where events $\xi_{1},\;\xi_{2},\,\xi_{3}$ and $\xi_{4}$ defined
as follows: 
\begin{enumerate}
\item Event $\xi_{1}$ occurs when $\Vert H(\ell)\Vert^{2}>\tau$ and $\Vert H_{c}(\ell)\Vert^{2}>\tau$. The 
resulting joint channel for joint detection of CCWS into $\mathcal{H}_{1}(\ell)=[H^{T}(\ell)\;\; H_{c}^{T}(\ell)]^{T}$,
where $H(\ell)$ and $H_{c}(\ell)$ are i.i.d. channel realization
with Gaussian distribution of zero mean and unit variance. The channel
norm $\Vert\mathcal{H}_{1}(\ell)\Vert^{2}=\Vert H(\ell)\Vert^{2}+\Vert H_{c}(\ell)\Vert^{2}$
has Chi-square distribution. 
\item Event $\xi_{2}$ occurs when $\Vert H(\ell)\Vert^{2}\le\tau \mbox{ and }\Vert H_{c}(\ell)\Vert^{2}>\tau$.
The resulting joint channel response for joint detection of CCWS into
$\mathcal{H}_{2}(\ell)=[H^{T}(\ell)\;\; H_{s}^{T}(\ell)\;\; H_{c}^{T}(\ell)]^{T}$. 
\item Event $\xi_{3}$ occurs when $\Vert H(\ell)\Vert^{2}>\tau \mbox{ and }\Vert H_{c}(\ell)\Vert^{2}\le\tau$.
The resulting joint channel response for joint detection of CCWS into
$\mathcal{H}_{3}(\ell)=[H^{T}(\ell)\;\; H_{c}^{T}(\ell)\;\; H_{sc}^{T}(\ell)]^{T}$, where $H_{sc}(\ell)$ is channel gain
of the $\ell$-th subcarrier selected for retransmission during full retransmission of failed packet at data link layer.  
\item Event $\xi_{4}$ occurs when $\Vert H(\ell)\Vert^{2}\le\tau$ and
$\Vert H_{c}(\ell)\Vert^{2}\le\tau$. The resulting joint channel for joint
detection of CCWS into $\mathcal{H}_{4}(\ell)=[H^{T}(\ell)\;\; H_{s}^{T}(\ell)\;\; H_{c}^{T}(\ell)\;\; H_{sc}^{T}(\ell)]^{T}$,
where $H_{s}(\ell)$ and $H_{sc}(\ell)$ are the channels corresponding
to the selective channels from the first transmission and full retransmission
of CCWS method. Note that random variables $\Vert H_{s}(\ell)\Vert^{2}$
and $\Vert H_{sc}(\ell)\Vert^{2}$ are i.i.ds. with Chi-square distribution
of degree  $2n_{r}$ each. 
\end{enumerate}
Second and third terms in \eqref{eq:CCWSjoitBER} are equivalent due
to the fact that $P(\xi_{2})=P(\xi_{3})$ and random variables $\Vert\mathcal{H}_{2}(\ell)\Vert$ and
$\Vert\mathcal{H}_{3}(\ell)\Vert$ are i.i.ds. Therefore, $E_{H}\Big[P(\xi_{2})P_{e\vert\xi_{2}}+P(\xi_{3})P_{e\vert\xi_{3}}\Big]=2E_{H}\Big[P(\xi_{2})P_{e\vert\xi_{2}}\Big]$.
Note that all channel realizations of the $\ell$-th subcarrier of
an OFDM system are i.i.d. with Gaussian distribution of mean zero and
unit variance. In order to achieve joint upper bound on BER for joint detection of CCWS method, we rewrite \eqref{eq:CCWSjoitBER} 
as follows:
\begin{align}
 & P_{e_{2}}=cE_{H}\Bigg[P(\xi_{1})Q\left(\sqrt{\frac{g \Vert \;\mathcal{H}_{1}(\ell)\Vert^{2}}{N_{0}}}\right)+2P(\xi_{2})\cdot\nonumber \\
\negmedspace\!\!\!\!\!\!\!\!\!\! & Q\left(\negmedspace\negmedspace\sqrt{\frac{g\Vert \;\mathcal{H}_{2}(\ell)\Vert^{2}}{N_{0}}}\negmedspace\right)\negmedspace\negmedspace+\negmedspace P(\xi_{4})Q\left(\negmedspace\negmedspace\sqrt{\frac{g\Vert \;\mathcal{H}_{4}(\ell)\Vert^{2}}{N_{0}}}\right)\negmedspace\negmedspace\Bigg].
\label{eq:CCWSJointQfunc}
\end{align}
Note that $\Vert \mathcal{H}_{1}(\ell)\Vert^{2}=\chi+\chi_{c}$ in the first term of \eqref{eq:CCWSJointQfunc} is sum if two i.i.d. chi-square random variables , where $\chi>\tau$ and  $\chi_{c}> \tau$.  Using approximation of Q-function in \cite{QFuncApprox} and, following \eqref{eq:SCC2} and \eqref{eq:SCC3}, we have
\begin{align}
\begin{split}
E_{H}\left[P(\xi_{1})P_{e}\vert\xi_{1}\right]  =  &cE_{H}\left[P(\xi_{1})Q\left(\sqrt{\frac{g\Vert\mathcal{H}_{1}(\ell)\Vert^{2}}{N_{0}}}\right)\right]\\
\le & \frac{c}{12}\int_{\tau}^{\infty}\exp(-g\frac{x}{2N_{0}})f_{X}(x)dx.\int_{\tau}^{\infty}\exp(-g\frac{x_{c}}{2N_{0}})f_{X_{c}}(x_{c})dx_{c}+\\
  &\frac{c}{4}\int_{\tau}^{\infty}\exp(-g\frac{4x}{3.2N_{0}})f_X(x).dx\int_{\tau}^{\infty}\exp(-g\frac{4x_{c}}{3.2N_{0}})f_{X_{c}}(x_{c})dx_{c}\\
= & \frac{c}{12}\left(\frac{\rho}{\sigma}\right)^{4n_{r}}\left(\exp\left(-\frac{\tau}{2\rho^{2}}\right)\sum_{k=0}^{n_{r}-1}\frac{1}{k!}\left(\frac{\tau}{2\rho^{2}}\right)^{k}\right)^{2}+\\
 & \frac{c}{4}\left(\frac{\rho_{1}}{\sigma}\right)^{4n_{r}}\left(\exp\left(-\frac{\tau}{2\rho_{1}^{2}}\right)\sum_{k=0}^{n_{r}-1}\frac{1}{k!}\left(\frac{\tau}{2\rho_{1}^{2}}\right)^{k}\right)^{2}.
\label{eq:CCWS1}
\end{split}
\end{align}
Similarly, 
\begin{align}
\begin{split}
2E_{H}&\left[P(\xi_{2})P_{e}\vert\xi_{2}\right]  =
  2cE_{H}\left[P(\xi_{2})Q\left(\sqrt{\frac{g\Vert\mathcal{H}_{2}(\ell)\Vert^{2}}{N_{0}}}\right)\right]\\
\le & \frac{c}{6}\int_{0}^{\tau}\exp(-g\frac{x}{2N_{0}})f_{X}(x)dx.\int_{0}^{\infty}\exp(-g\frac{x_{s}}{2N_{0}})f_{X_{s}}(x)dx_{s}\int_{\tau}^{\infty}\exp(-g\frac{x_{c}}{2N_{0}})f_{X_{c}}(x_{c})dx_{c}+\\
 & \frac{c}{2}\int_{0}^{\tau}\exp(-g\frac{4x}{3.2N_{0}})f_{X}(x)dx.\int_{0}^{\infty}\exp(-g\frac{4x_{s}}{3.2N_{0}})f_{X_{s}}(x)dx_{s}\int_{\tau}^{\infty}\exp(-g\frac{4x_{c}}{3.2N_{0}})f_{X_{c}}(x_{c})dx_{c},
\end{split}
\label{eq:SubEqCCWS}
\end{align}
where $\chi<\tau$,   $\chi_s \in \mathcal{R}$ and  $\chi_{c}> \tau$. Simplifying \eqref{eq:SubEqCCWS}, we have 
\begin{align}
\begin{split}
2E_{H}&\left[P(\xi_{2})P_{e}\vert\xi_{2}\right]  \le  \frac{c}{6}\left(\frac{\rho}{\sigma}\right)^{6n_{r}}\left(1-\exp\left(-\frac{\tau}{2\rho^{2}}\right)\sum_{k=0}^{n_{r}-1}\frac{1}{k!}\left(\frac{\tau}{2\rho^{2}}\right)^{k}\right)\left(\exp\left(-\frac{\tau}{2\rho^{2}}\right)\sum_{k=0}^{n_{r}-1}\frac{1}{k!}\left(\frac{\tau}{2\rho^{2}}\right)^{k}\right)+\\
 & \frac{c}{2}\left(\frac{\rho_{1}}{\sigma}\right)^{6n_{r}}\left(1-\exp\left(-\frac{\tau}{2\rho_{1}^{2}}\right)\sum_{k=0}^{n_{r}-1}\frac{1}{k!}\left(\frac{\tau}{2\rho_{1}^{2}}\right)^{k}\right)\left(\exp\left(-\frac{\tau}{2\rho_{1}^{2}}\right)\sum_{k=0}^{n_{r}-1}\frac{1}{k!}\left(\frac{\tau}{2\rho_{1}^{2}}\right)^{k}\right).
\label{eq:CCWS2}
\end{split}
\end{align}
Also, it can be shown that
\begin{align}
\begin{split}
E_{H}&\left[P(\xi_{4})P_{e}\vert\xi_{4}\right]  =
  cE_{H}\left[P(\xi_{4})Q\left(\sqrt{\frac{g\Vert\mathcal{H}_{4}(\ell)\Vert^{2}}{N_{0}}}\right)\right]\\
\le & \frac{c}{12}\Big(\frac{\rho}{\sigma_{h}}\Big)^{8n_{r}}\Bigg(1-\exp\Big(-\frac{\tau}{2\rho^{2}}\Big)\sum_{k=0}^{n_{r}-1}\frac{1}{k!}\Big(\frac{\tau}{2\rho^{2}}\Big)^{k}\Bigg)^{2}+\\
 & \frac{c}{4}\Big(\frac{\rho_{1}}{\sigma_{h}}\Big)^{8n_{r}}\Bigg(1-\exp\Big(-\frac{\tau}{2\rho_{1}^{2}}\Big)\sum_{k=0}^{n_{r}-1}\frac{1}{k!}\Big(\frac{\tau}{2\rho_{1}^{2}}\Big)^{k}\Bigg)^{2},
\label{eq:CCWS3}
\end{split}
\end{align}
where $\chi < \tau$, $\chi_s\in \mathcal{R}$, $\chi_c < \tau$ and $\chi_{sc}\in \mathcal {R}$.
Now  using \eqref{eq:CCWS1},  \eqref{eq:CCWS2} and \eqref{eq:CCWS3} in \eqref{eq:CCWSJointQfunc}, we have
\begin{align}
\begin{split}
P_{e_{2}} & \le\frac{c}{12}\Big(\frac{\rho}{\sigma}\Big)^{4n_{r}}\Bigg(\exp\Big(-\frac{\tau}{2\rho^{2}}\Big)\sum_{k=0}^{n_{r}-1}\frac{1}{k!}\Big(\frac{\tau}{2\rho^{2}}\Big)^{k}\Bigg)^{2}+\frac{c}{4}\Big(\frac{\rho_{1}}{\sigma}\Big)^{4n_{r}}\Bigg(\exp\Big(-\frac{\tau}{2\rho_{1}^{2}}\Big)\sum_{k=0}^{n_{r}-1}\frac{1}{k!}\Big(\frac{\tau}{2\rho_{1}^{2}}\Big)^{k}\Bigg)^{2}+\\
&\frac{c}{6}\Big(\frac{\rho}{\sigma}\Big)^{6n_{r}} \Bigg(\exp\Big(-\frac{\tau}{2\rho^{2}}\Big)\sum_{k=0}^{n_{r}-1}\frac{1}{k!}\Big(\frac{\tau}{2\rho^{2}}\Big)^{k}\Bigg)\Bigg(1-\exp\Big(-\frac{\tau}{2\rho^{2}}\Big)\sum_{k=0}^{n_{r}-1}\frac{1}{k!}\Big(\frac{\tau}{2\rho^{2}}\Big)^{k}\Bigg)+\\
&\frac{c}{2}\Big(\frac{\rho_{1}}{\sigma}\Big)^{6n_{r}}\Bigg(\exp\Big(-\frac{\tau}{2\rho_{1}^{2}}\Big)\sum_{k=0}^{n_{r}-1}\frac{1}{k!}\Big(\frac{\tau}{2\rho_{1}^{2}}\Big)^{k}\Bigg) \Bigg(1-\exp\Big(-\frac{\tau}{2\rho_{1}^{2}}\Big)\sum_{k=0}^{n_{r}-1}\frac{1}{k!}\Big(\frac{\tau}{2\rho_{1}^{2}}\Big)^{k}\Bigg)+\\
&\frac{c}{12}\Big(\frac{\rho}{\sigma_{h}}\Big)^{8n_{r}}\Bigg(1-\exp\Big(-\frac{\tau}{2\rho^{2}}\Big)\sum_{k=0}^{n_{r}-1}\frac{1}{k!}\Big(\frac{\tau}{2\rho^{2}}\Big)^{k}\Bigg)^{2}\\
 & +\frac{c}{4}\Big(\frac{\rho_{1}}{\sigma_{h}}\Big)^{8n_{r}}\Bigg(1-\exp\Big(-\frac{\tau}{2\rho_{1}^{2}}\Big)\sum_{k=0}^{n_{r}-1}\frac{1}{k!}\Big(\frac{\tau}{2\rho_{1}^{2}}\Big)^{k}\Bigg)^{2}.
\label{eq:BERUBoundJointCCWS}
\end{split}
\end{align}
\noindent

In next section, we present throughput analysis and optimization with
respect parameter $\tau$ for SCC and CCWS.

\section{Throughput Analysis}

\label{sec:throughAnalysis} Now we present throughput analysis of the proposed SCC and CCWS methods. In throughput analysis, we consider infinite
many rounds of retransmission. One round of retransmission consists
of first transmission followed by a retransmission. The receiver preserves
observation corresponding to the first transmission for joint detection.
We consider joint detection for retransmission by combining observation
from the first transmission and subsequent \emph{partial or full}
retransmission. In practice, transceiver pair continues retransmission
rounds until error free packet is received or maximum number of retransmission
rounds are reached. For throughput analysis, we follow conventional definition of throughput
$\eta$, which is ratio of error-free information bits received $k$
to the total number of bits transmitted $n$ ( $\eta=\frac{k}{n}$).
Let $P_{e}$ and $P_{e_{s}}$ be the bit-error probabilities of the first
transmission and joint detection followed by retransmission, respectively.
Assuming that each bit in the frame is independent, probability of
receiving an error-free packet of length $L_{f}$ with probability
of bit-error $P_{e}$ is $p_{c}=(1-P_{e})^{L_{f}}$. Then probability
of receiving a bad packet is $p_{\epsilon}=1-p_{c}$. Similarly, for joint detection, $p_{c_{s}=(1-P_{e_s}})^{L_{f}}$
and $p_{\epsilon_{s}}=1-p_{c_{s}}$. Next, we present throughput analysis
of SCC method.
\subsection{Throughput analysis of SCC}

\label{subsec:throughAnalsisSCC} 
Now we provide throughput analysis
of SCC method. In response to the NACK signal in the event of packet failure, receiver initiates selective retransmission of 
poor quality subcarriers instead of full retransmission. Let $m$ be the fraction of subcarriers selected for retransmission. That is, at a given SNR, fraction $m=P(\Vert H(\ell)\Vert^2\le \tau)$.
In SCC, each retransmission round consists of first transmission followed by
\emph{partial} retransmission resulting in to transmission of $k(1+m)$ bits in each round. Let $p_{c}$ and $p_{c_{s}}$ be the
probabilities of receiving correct frame from the first transmission
and joint detection of SCC method, respectively. The probabilities
of receiving bad frame from the first transmission and joint detection
of SCC are $p_{\epsilon}$ and $p_{\epsilon_{s}}$, respectively, for SCC. The probabilities of frame error $p_{\epsilon}$ and $p_{\epsilon_{s}}$ are functions of BER given in \eqref{eq:BERUpperBoundJointSCC} and
\eqref{eq:BERUpperBoundSingleSCC}, respectively. Thus, expected number of information bits $n$ to be transmitted
to receive $k$ error-free bits are \cite{ShuCostMiller}
\begin{align}
\begin{split}
  n=&kp_{c}+k(1+m)p_{\epsilon}p_{c_{s}}+k(2+m)p_{\epsilon}p_{\epsilon_{s}}p_{c}\negthinspace\negthinspace+k(2+2m)p_{\epsilon}^{2}\\
 & p_{\epsilon_{s}}p_{c_{s}}\negthinspace\negthinspace+k(3+2m)p_{\epsilon}^{2}p_{\epsilon_{s}}^{2}p_{c}+k(3+3m)p_{\epsilon}^{3}p_{\epsilon_{s}}^{2}p_{c_{s}}\negthinspace\negthinspace+k(4+3m)\\
 & p_{\epsilon}^{3}p_{\epsilon_{s}}^{3}p_{c}+k(4+4m)p_{\epsilon}^{4}p_{\epsilon_{s}}^{3}p_{c_{s}}+k(5+4m)p_{\epsilon}^{4}p_{\epsilon_{s}}^{4}p_{c}+\cdots
\end{split}
\label{eq:ThputSCC1}
\end{align}
Now rearranging \eqref{eq:ThputSCC1}, we have
\begin{align}
 \begin{split}
 n=&kp_{c}\Big(1+(2+m)p_{\epsilon}p_{\epsilon_{s}}+(3+2m)p_{\epsilon}^{2}p_{\epsilon_{s}}^{2}+(4+3m) \\
 & p_{\epsilon}^{3}p_{\epsilon_{s}}^{3}\negthinspace\negthinspace\negthinspace\negthinspace+(5+4m)p_{\epsilon}^{4}p_{\epsilon_{s}}^{4}+\cdots\Big)\negthinspace\negthinspace+kp_{\epsilon}p_{c_{s}}\Big((1+m)+(2+2m) \\
 & p_{\epsilon}p_{\epsilon_{s}}+(3+3m)p_{\epsilon}^{2}p_{\epsilon_{s}}^{2}+(4+4m)p_{\epsilon}^{3}p_{\epsilon_{s}}^{3}+\cdots\Big) \\
 =&kp_{c}\Big(1+(2+m)\alpha+(3+2m)\alpha^{2}+(4+3m)\alpha^{3}+\hdots\Big) \\
 & +kp_{\epsilon}p_{c_{s}}\Big((1+m)+(2+2m)\alpha+(3+3m)\alpha^{2}+ \\
 & (4+4m)\alpha^{3}+(5+5m)\alpha^{4}+\cdots\Big) \\
=&k.p_{c}(1+m\alpha)(1+2\alpha+3\alpha^{2}+4\alpha^{3}+\cdots) \\
&+kp_{\epsilon}p_{c_{s}}(1+m) 
 \Big(1+2\alpha+3\alpha^{2}+4\alpha^{3}+\cdots\Big)
\end{split}
\label{eq:ThputSCC2}
\end{align}
where $\alpha=p_{\epsilon}p_{\epsilon_{s}}$. Note that $1+2\alpha+3\alpha^{2}+4\alpha^{3}+\cdots=\frac{1}{(1-\alpha)^{2}}$
\cite{bromwich1908introduction}. Therefore,
\begin{equation}
n=\frac{kp_{c}(1+m\alpha)+p_{\epsilon}p_{c_{s}}(1+m)}{(1-\alpha)^{2}}.
\label{eq:Totalbits}
\end{equation}
Thus, throughput of SCC methods is 
\begin{align}
\eta_s= & \dfrac{k}{n}=\dfrac{(1-\alpha)^{2}}{p_{c}(1+m\alpha)+p_{\epsilon}p_{c_{s}}(1+m)}.
\label{Eq:ThroughputSCC}
\end{align}
It is clear from \eqref{eq:BERUpperBoundJointSCC}, \eqref{eq:BERUpperBoundSingleSCC} and \eqref{Eq:ThroughputSCC}
that throughput $\eta_s=f(\tau, SNR)$. Next, we present throughput analysis
of CCWS method.
\subsection{Throughput analysis of CCWS}
\label{subsec:throughAnalsisCCWS} In CCWS method, modulation layer
executes selective retransmission for first transmission of a packet and full retransmission
in the event of CRC failure. Thus, for each of the first transmission and subsequent
full retransmission of OFDM symbol initiated by Chase combining protocol
at data-link layer, $k(1+m)$ information bits are transmitted,
where additional $km$ bits are the results of selective retransmission
for the first transmission and full retransmission. Similar to SCC
method, $P_{e_1}$ and $P_{e_{2}}$ be the probability of error for
the first transmission and joint detection corresponding to the retransmission. Since CCWS method employs selective Chase combining during first transmission, the bit-error probability of the first transmission is given in \eqref{eq:BERUpperBoundJointSCC}.
The bit-error rate $P_{e_{2}}$ for joint detection
CCWS is given in \eqref{eq:BERUBoundJointCCWS},
respectively. We denote  probabilities of receiving error-free frame of length $L_f$ for the first transmission and 
selective retransmission by $p_{c_1}$ and $p_{c_2}$, respectively. The corresponding probabilities of receiving incorrect frame for first transmission and joint detection in the event of packet failure are $p_{\epsilon_1} =1-p_{c_1}$ and $p_{\epsilon_2} =1-p_{c_2}$, respectively. The expected number of information bits transmitted
to deliver error-free $k$ information bits for CCWS method are \cite{ShuCostMiller}
\begin{align}
\begin{split}
n= & k.(1+m)p_{c_1}+2k(1+m)p_{\epsilon_1}.p_{c_{2}}+k3(1+m)p_{\epsilon_1}p_{\epsilon_{2}}p_{c_1}+ \\
 & 4k(1+m)p_{\epsilon_1}^{2}p_{\epsilon_{2}}p_{c_{2}}+k.5(1+m)p_{\epsilon_1}^{2}p_{\epsilon_{2}}^{2}p_{c_1}+6k(1+m) \\
 & p_{\epsilon_1}^{3}p_{\epsilon_{2}}^{2}p_{c_{2}}+k.7(1+m)p_{\epsilon_1}^{3}p_{\epsilon_{2}}^{3}p_{c_1}+\hdots. \\
= & k.(1+m)p_{c_1}\Big(1+3p_{\epsilon_1}p_{\epsilon_{2}}+5p_{\epsilon_1}^{2}p_{\epsilon_{2}}^{2}+\hdots\Big)+ \\
 &2k(  1+m)p_{\epsilon_1}p_{c_{2}}\Big(\!1+2p_{\epsilon_1}p_{\epsilon_{2}}\!+\!\!3p_{\epsilon_1}^{2}p_{\epsilon_{2}}^{2}\!+\hdots\Big).
\label{eq:throughputexpressionJointatPhysical}
\end{split}
\end{align}
Let $\alpha=p_{\epsilon_1}p_{\epsilon_{2}}$, then 
\begin{align}
\begin{split}
n  =&k(1+m)p_{c_1}\Big(1+3\alpha+5\alpha^{2}+7\alpha^{3}+\hdots\Big)+ \\
 & 2k(1+m)p_{\epsilon_1}.p_{c_{2}}\Big(1+2\alpha+3\alpha^{2}+4\alpha^{3}+\hdots\Big).
\end{split}
\label{eq:totalBitsCCWS}
\end{align}
After simplification of \eqref{eq:totalBitsCCWS}, we have \cite{bromwich1908introduction}
\begin{equation}
n=k(1+m)p_{c_1}\frac{1+\alpha}{(1-\alpha)^{2}}+2k(1+m)p_{\epsilon}p_{c_{2}}\frac{1}{(1-\alpha)^{2}}.
\end{equation}
Throughput of CCWS method is 
\begin{align}
\eta_{c} & =\frac{k}{n}=\frac{(1-\alpha)^{2}}{\Big(p_{c_1}(1+\alpha)+2(1-p_{c_1})p_{c_{2}}\Big)\quad(1+m)}.
\label{eq:throtputCCWS}
\end{align}
Note that similar to the throughput $\eta_s$ for SCC method, throughput $\eta_{c}$ of CCWS 
is also function  of parameter $\tau$ that controls the information to
be transmitted during selective retransmission. The parameter $\tau$
can be optimize to maximize throughput under OFDM signaling. Next,
we discuss search for optimal $\tau$ for SCC and CCWS to enhance
throughput of OFDM transceiver.

\section{Throughput Optimization}

\label{sec:OptThrough} 
\begin{figure}[t]
\centering \includegraphics[width=10cm]{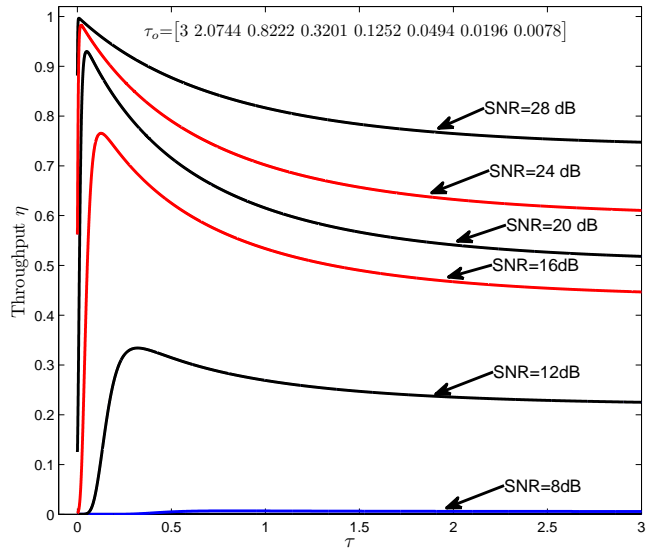}
\caption{Optimal $\tau_o$ for different SNR operating point that maximizes analytical throughput of SCC method in \eqref{Eq:ThroughputSCC}.}
\label{fig:OptimatauforSCC} 
\end{figure}

\begin{figure}[t]
\centering{}\includegraphics[width=10cm]{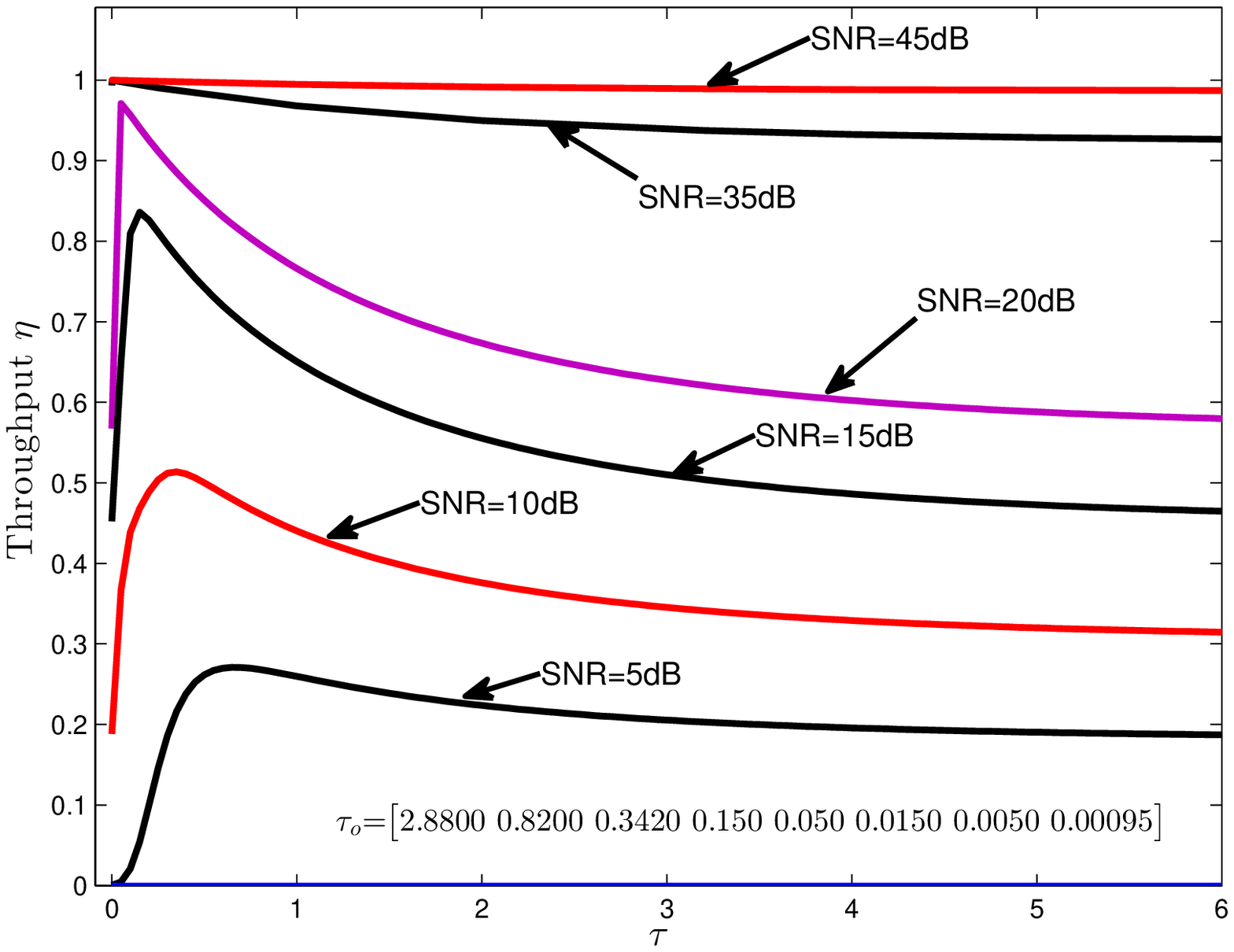}
\caption{Optimal $\tau_o$ for different SNR operating point that maximizes analytical throughput of CCWS method in \eqref{eq:throtputCCWS}.}
\label{fig:TauVsThroughputCCWS} 
\end{figure}

In this section, we optimize throughput
of the proposed selective retransmission methods at modulation layer. The amount
of information that a receiver request to the transmitter 
in the event of a packet failure has direct impact on the throughput
of the transceiver. Most of the time, especially in high SNR regime, receiver
can recover from bit errors by receiving little more information and employing joint detection.
In selective retransmission at modulation level, threshold $\tau$
on channel norm of a subcarrier is measure of quality of a channel.
By choosing proper threshold $\tau$, receiver can request minimum
information needed to recover from errors for a failed packet. The threshold
$\tau$ is function of SNR and modulations such as 4-QAM and 16-QAM.
It is clear from \eqref{Eq:ThroughputSCC} and \eqref{eq:throtputCCWS}
that throughput of SCC and CCWS methods, respectively, are function
of frame-error rate (FER). Furthermore, FER is not a linear or quadratic
function of SNR and parameter $\tau$. Now we write unconstrained
optimization problem for throughput $\eta$ with respect to parameter
$\tau$ as follows:
\begin{equation}
\mathbf{\tau_{0}}=\underset{\tau}{\text{arg  max}}\quad \eta= f(\tau,SNR).
\label{eq:etao}
\end{equation}
Since throughput $\eta$ is non-convex function in parameter $\tau$,
optimal $\tau$ that maximizes throughput $\eta$ for each SNR can
be computed off-line using exhaustive search. Thus, a table of optimal
threshold $\tau$ to maximize throughput for  SNR operating points
can be generated using throughput expression in \eqref{Eq:ThroughputSCC} and \eqref{eq:throtputCCWS}
 for SCC and CCWS methods, respectively. Note that each retransmission
method has different throughput function $\eta=f(\tau,\, SNR)$. For
example, analytical throughput function for SCC and CCWS is given
in \eqref{Eq:ThroughputSCC} and \eqref{eq:throtputCCWS}, respectively.
In Section~\ref{Sec:simulation}, we maximize $\eta=f(\tau,\, SNR)$
with respect to parameter $\tau$. Note that parameter $\tau$ appears
in probability of frame error $p_{\epsilon}$, which is function of
probability of bit-error presented in Section~\eqref{sec:perfomnceanlysis}. Although $\eta=f(\tau,\, SNR)$ is
not linear or quadratic function of $\tau$ at a given SNR, Figure~\ref{fig:OptimatauforSCC} shows that for a given SNR, $\tau_o$ that maximizes throughput $\eta$ is unique for SCC method. The optimal $\tau_o$ can be computed off-line from throughput lower bound for SCC and CCWS using \eqref{Eq:ThroughputSCC} and \eqref{eq:throtputCCWS}, respectively. Based on channel condition, amount of information to be transmitted can be controlled using vector $\tau_o$. Figure~\ref{fig:TauVsThroughputCCWS} shows optimal threshold $\tau_o$ for  SNR points that maximizes throughput CCWS method. In low SNR regime, throughput $\eta$ is more sensitive to threshold $\tau$ as compared to high SNR regime. This is due to the fact that in high SNR regime, very few errors occur during first transmission resulting in to fewer retransmissions. We use threshold vector $\tau_o$ in next section to compute throughput.



\section{Simulation}
\label{Sec:simulation} 

\begin{figure}[t]
\centering \includegraphics[width=10cm]{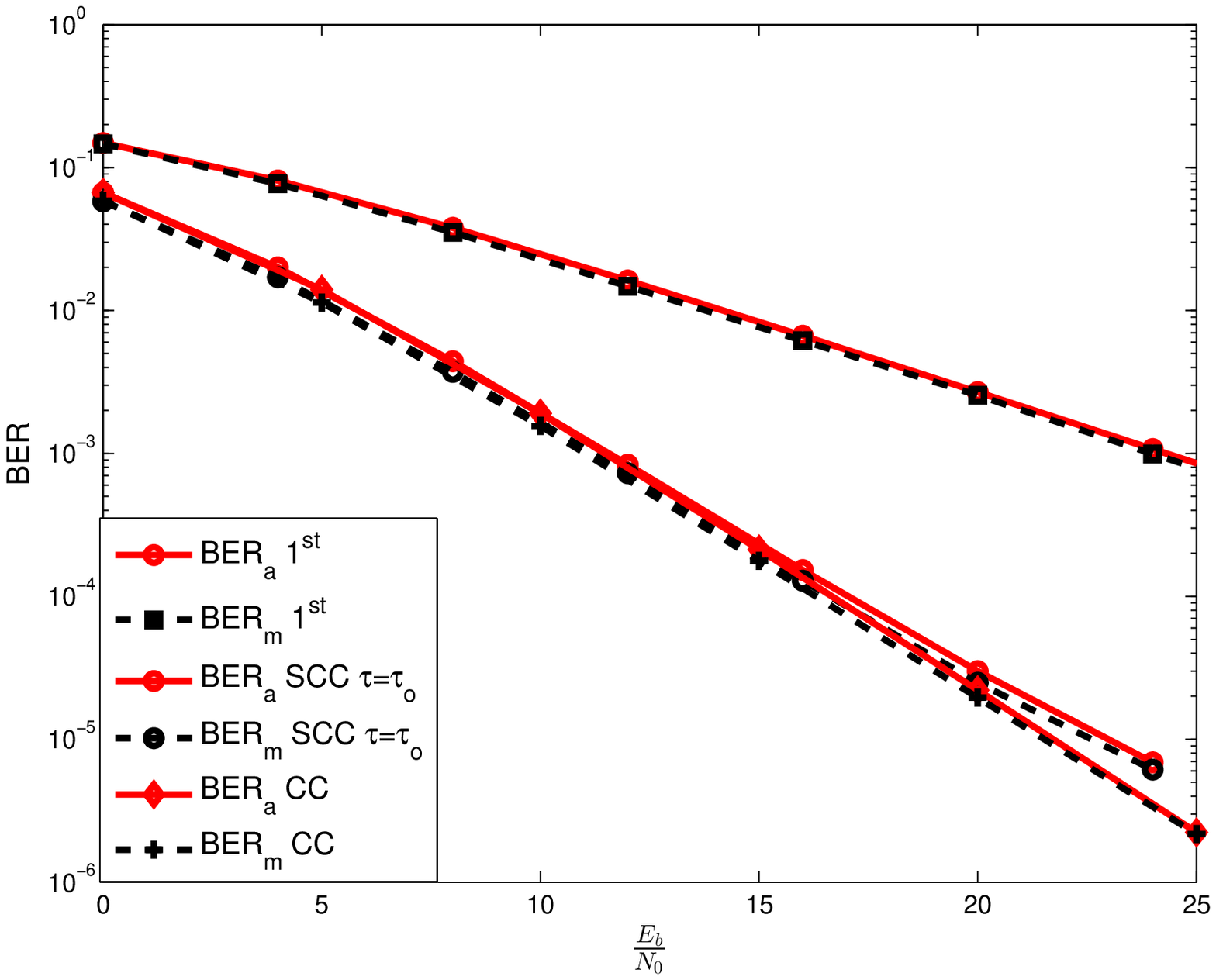}
 \caption{BER performance comparison of analytical bit-error rate in \eqref{eq:BERUpperBoundJointSCC} and Monte Carlo simulation for SCC method  of SISO-OFDM system. In analytical and simulation results, parameter $\tau=0$, $\tau_f$ and $\tau_o$ are used to target BER without selective retransmission(single), full retransmission and optimize throughput, respectively, for SCC method.}
\label{fig:BERAnalVsSimSCC} 
\end{figure}

Now we present efficiency for aforementioned SCC, CCWS and MSCC methods for optimized threshold $\tau_o$  on the throughput of OFDM systems. In simulation
setup, we consider 4-QAM constellation for OFDM
signaling under Rayleigh fading frequency selective channel. We use
OFDM signaling with $N_{s}=512$ subcarriers for 10-tap frequency
selective channel. Each complex OFDM channel realization has Gaussian
distribution with zero-mean and unit variance ($\sigma^2_h=1$). We
assume block fading channel in quasi-static fashion such that channel
remains highly correlated during transmission of one OFDM symbol. First,
we present comparison of  analytical BER upper
bound and BER from Monte Carlo simulation denoted by $BER_a$ and $BER_m$, respectively. We also provide throughput
results of SCC and CCWS methods in comparison with conventional Chase
combining method. We denote analytical and simulation throughput by $\eta_a$ and $\eta_m$, respectively. In order to maximize throughput, threshold $\tau$ on channel norm of OFDM
subcarriers is optimized for each SNR point of SCC and CCWS protocols. We analytically compute
threshold vector $\tau_o$ off-line to maximize throughput of SCC and CCWS methods from the analytical throughput using \eqref{Eq:ThroughputSCC} and \eqref{eq:throtputCCWS}, respectively. We also demonstrate that throughput gain our proposed CCWS and MSCC methods hold under CC-HARQ method at MAC layer. We consider half-rate LDPC code (648, 324) to evaluate efficacy of CCWS and MSCC methods. We denote CCWS and MSCC methods with FEC enabled by CCWS-HARQ and MSCC-HARQ in simulation results.  
 
\begin{figure}[ht]
\centering \includegraphics[width=10cm]{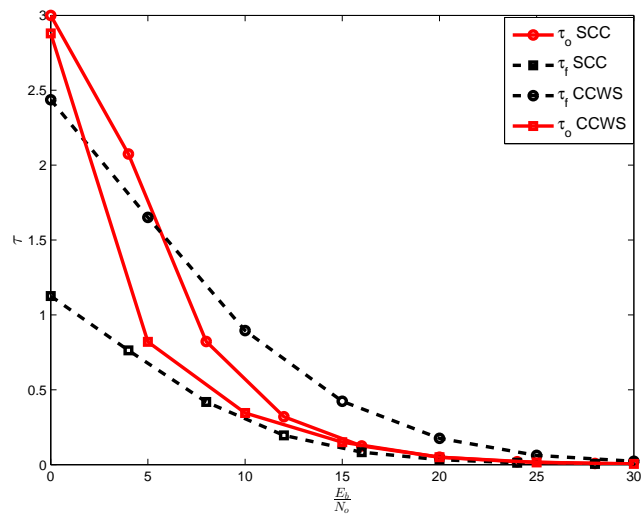}
 \caption{Threshold comparison $\tau_f$ and $\tau_o$ corresponding to target BER equals 
full retransmission and optimal throughput for both  SCC and CCWS   protocols, respectively.} 
\label{fig:tauComparison1} 
\end{figure}

\begin{figure}[ht]
\centering \includegraphics[width=9cm]{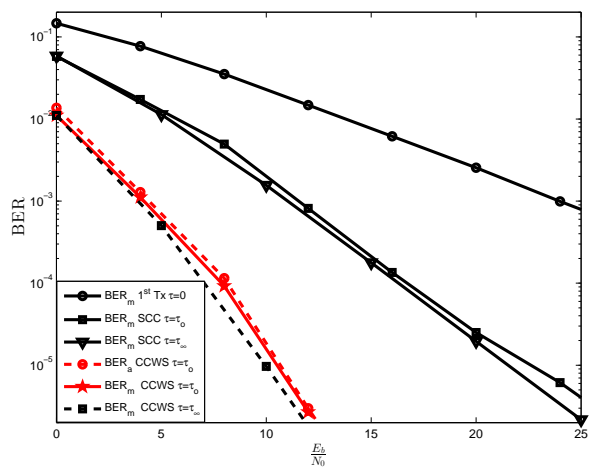}
 \caption{BER performance comparison of analytical bit-error rate in \eqref{eq:BERUBoundJointCCWS} and Monte Carlo simulation for CCWS method  of SISO-OFDM system. In analytical and simulation results, parameter $\tau=0,\tau_f$ and $\tau_o$ are used to target BER corresponding to full retransmission and  optimize throughput for both first transmission and retransmission, respectively, for CCWS method.}
\label{fig:BERAnalVssimuCCWS} 
\end{figure}

First, we compare BER performance for single (first) transmission and  subsequent retransmission in the event of failed packet. In particular, we compare BER of full retransmission (Chase combining) and SCC in Figure~\ref{fig:BERAnalVsSimSCC}. For analytical and Monte Carlo runs of SCC, we use  optimal threshold vector $\tau_o$  that maximizes throughput. In Figure \ref{fig:BERAnalVsSimSCC}, we  present BER comparison of upper bound and BER from Monte Carlo simulation of SCC under SISO-OFDM system using optimal threshold vector $\tau_o$. 
Note that BER performance of SCC with threshold $\tau_o$ and $\tau_f$ is similar, where threshold $\tau_f$ targets BER of SCC equals full retransmission. 
Figure~\ref{fig:tauComparison1} compares threshold vector $\tau_o$  that optimizes throughput of SCC and CCWS at different SNR points and threshold vector $\tau_f$ to target BER of SCC and CCWS equals the corresponding full retransmission. Difference between vectors $\tau_o$  and $\tau_f$  is marginal at moderate and high SNR. At high SNR, BER of SCC with threshold $\tau_o$ is marginally  higher than that of BER for threshold $\tau_f$ (BER of full retransmission). It is clear from Figure \ref{fig:BERAnalVsSimSCC} that  retransmission of partial information corresponding to poor quality subcarriers achieves BER of full retransmission. The threshold $\tau$ on channel norm of subcarriers controls amount of partial information to be retransmitted.
  Figure~\ref{fig:BERAnalVssimuCCWS} compares BER performance of  CCWS method with conventional CC method for different  values of threshold vector $\tau$ on channel norm of subcarriers.  First transmission of packet under CCWS method, which employs selective retransmission at modulation level, achieves BER of full retransmission with threshold vector $\tau_o$ as shown in Figure~\ref{fig:BERAnalVssimuCCWS}.  Threshold vector $\tau_o$ for CCWS method optimizes throughput of CCWS. The comparison of $\tau_o$ for SCC and CCWS is presented in  Figure~\ref{fig:tauComparison1}. Figure~\ref{fig:BERAnalVssimuCCWS} also reveals that using threshold $\tau_o$ that maximizes throughput of CCWS method, BER of joint detection under CCWS method is similar to the BER of four receive antennas. Note that threshold $\tau_{\infty}$ implies  retransmissions of all subcarriers and $\tau_{\infty}$ for CCWS in fact is equivalent to  four transmissions of a packets or four receive antennas. In Figure~\ref{fig:BERAnalVssimuCCWS}, we also compare analytical BER ($\text{BER}_a$) of joint detection  given in  \eqref{eq:BERUBoundJointCCWS} with Monte Carlo BER ($\text{BER}_m$) of CCWS for threshold vector $\tau_o$. It is clear from Figure~\ref{fig:BERAnalVsSimSCC} and Figure~\ref{fig:BERAnalVssimuCCWS} that \eqref{eq:BERUpperBoundJointSCC} and \eqref{eq:BERUBoundJointCCWS} are tight BER upper bounds for joint detections of SCC and CCWS  methods, respectively. In CCWS method, threshold $\tau_o$ is computed using \eqref{eq:etao} to maximize throughput $\eta$. Similar to SCC method, we compute threshold $\tau_f$ that targets BER of CCWS equals to the BER of joint detection of four transmissions. For CCWS method, $\mbox{BER}_m$ for threshold vector $\tau_f$ is lower than $\mbox{BER}_m$ for threshold vector $\tau_o$.  
\begin{figure}[ht]
\centering \includegraphics[width=9cm]{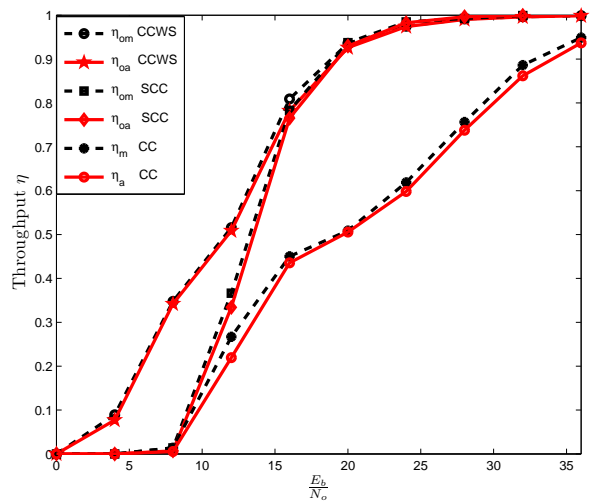} 
\caption{ Simulation and Analytical Throughput comparison for SCC, CCWS and CC using \eqref{Eq:ThroughputSCC} and \eqref{eq:throtputCCWS} for $\tau_o$.}.
\label{fig:througputCCWSptimizedVsConventionCC} 
\end{figure}

Throughput is the key performance metric of a communication system and throughput optimization of selective retransmission based on channel quality for OFDM system is the main focus of this work. For throughput computation, we use standard definition of  throughput of a communication system  \cite{ShuLin} as 
\begin{align}
\eta=\frac{\mbox{\,Error-free bits received}}{\mbox{Total bits transmitted}}=\dfrac{k}{n}.
\end{align}
 Now we provide throughput  results to demonstrate the efficacy of selective retransmission on spectral efficiency for SCC and CCWS methods at modulation layer. We use throughput gain of optimized $\tau_o$ for SCC and CCWS methods over conventional Chase combining method as a measure of spectral efficiency. Figure \ref{fig:througputCCWSptimizedVsConventionCC}
provides throughput comparison of conventional Chase combining, SCC and CCWS methods.  Monte Carlo throughput (simulation) $\eta_{om}$ for SCC and CCWS is computed using their respective optimal thresholds $\tau_o$. The throughput performance of both SCC and CCWS is much higher than conventional Chase combining method.  Throughput in Figure \ref{fig:througputCCWSptimizedVsConventionCC} also reveals that CCWS outperforms SCC in low SNR regime. This is due to the fact that at low SNR,  joint  detection using  more retransmission improves throughput of the communication systems. Figure \ref{fig:througputCCWSptimizedVsConventionCC} also compares analytical throughput $\eta_{oa}$ and simulation throughput $\eta_{om}$ of SCC and CCWS methods. Analytical throughput $\eta_{oa}$ of SCC and CCWS is computed using \eqref{Eq:ThroughputSCC} and \eqref{eq:throtputCCWS}, respectively. The throughput expressions for SCC and CCWS in   \eqref{Eq:ThroughputSCC} and \eqref{eq:throtputCCWS}, respectively, provide tight lower bounds on throughput as shown in Figure \ref{fig:througputCCWSptimizedVsConventionCC}.  

\begin{figure}[ht]
\centering \includegraphics[width=9cm]{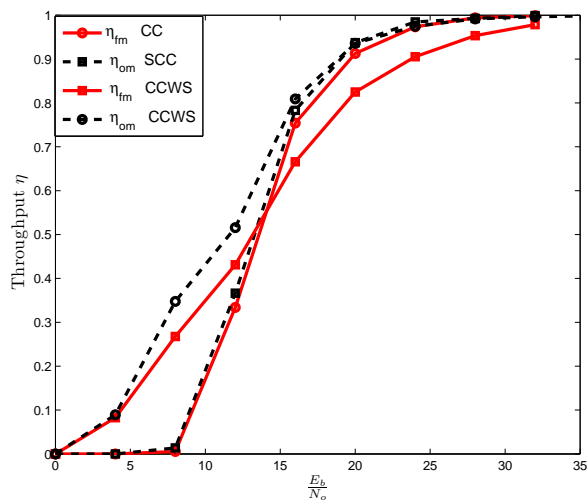}
\caption{Throughput comparison of Optimal SCC and CCWS using Target BER of Full retransmission $\tau=0$ Vs optimal
SCC form Figure \ref{fig:tauComparison1}.}.
\label{fig:SCCandCCWSforTaufTauo} 
\end{figure}

Now we present impact of thresholds $\tau_o$ and $\tau_f$ on the  throughput of SCC and CCWS methods. Figure \ref{fig:SCCandCCWSforTaufTauo} compares simulation results of throughput $\eta_{om}$ and $\eta_{fm}$ corresponding to the optimal threshold $\tau_o$ that maximizes throughput and threshold $\tau_f$ that achieves BER of full retransmission of the proposed methods (SCC and CCWS). The simulation results suggest that throughput of SCC method for $\eta_{om}$ and $\eta_{fm}$ are similar. Throughput for $\eta_{om}$ are marginally better than for $\eta_{fm}$.  However, throughput $\eta_{om}$ of CCWS with optimal threshold vector $\tau_o$ is higher than  throughput $\eta_{fm}$ of CCWS with threshold vector $\tau_f$ that achieves BER of full retransmission.
\begin{figure}[ht]
\centering \includegraphics[width=9cm]{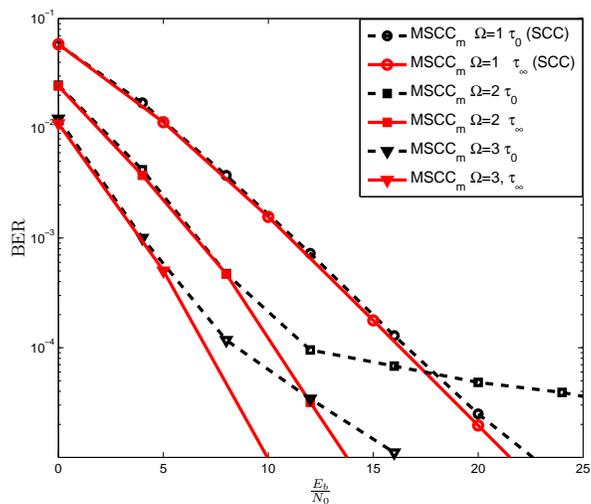}
\caption{BER comparison of  SCC and MSCC for $\tau_o$ and $\tau_\infty$.}.
\label{fig:BERMSCC} 
\end{figure}

\begin{figure}[ht]
\centering \includegraphics[width=9cm]{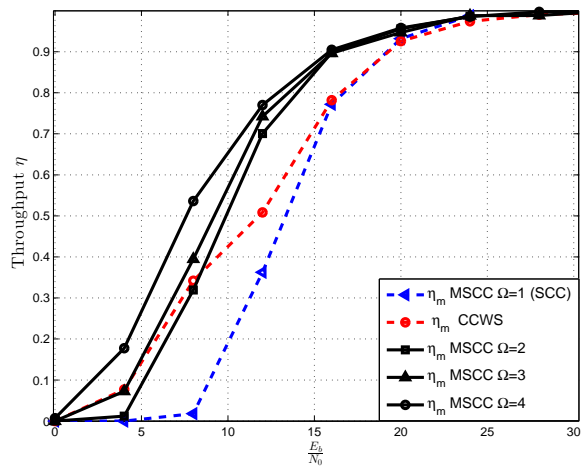}
\caption{Throughput comparison of  SCC, CCWS and MSCC.}.
\label{fig:MSCC} 
\end{figure}
\begin{figure}[ht]
\centering \includegraphics[width=9cm]{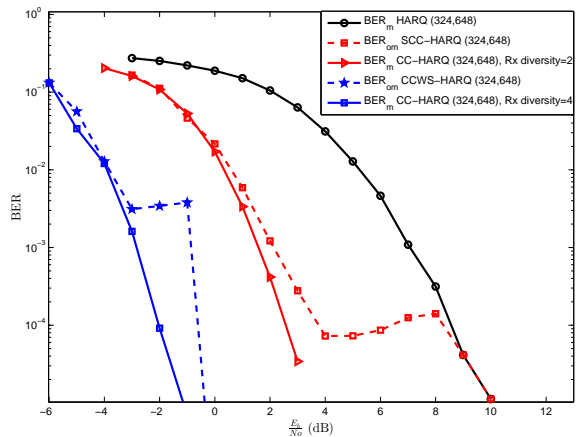}
\caption{BER comparison of ARQ, SCC, CCWS, CC Vs HARQ, SCC-HARQ, CCWS-HARQ and CC-HARQ at $\tau=0$, $\tau=\tau_o$ and $\tau=\infty$ respectively using half rate LDPC code.}.
\label{fig:HBERSCCwithFEC} 
\end{figure}

BER of MSCC for retransmission count $\Omega=1, 2 \mbox{ and } 3 $ for optimal threshold $\tau_0$ on channel norm  is presented in Figure~\ref{fig:BERMSCC}. The threshold values of vector $\tau_0$ that maximize throughput $\eta$ at corresponding SNR point are obtained from exhaustive search using \eqref{eq:etao}. Figure~\ref{fig:BERMSCC} reveals that lower BER does not result into higher throughput. The BER performance corresponding to  $\tau_\infty$ is lower than $\tau_0$. However, $\eta_o > \eta_\infty$.   Figure~\ref{fig:MSCC} presents throughput gain of MSCC method over SCC (MSCC with single retransmission) and CCWS method. It is clear from Figure~\ref{fig:MSCC} that for two retransmissions ($\Omega=2$), MSCC achieves significant throughput gain as compared to SCC.   Throughput of MSCC with joint detection of three selective retransmissions is higher than CCWS as well. As shown in Figure~\ref{fig:MSCC}, there is throughput improvement by jointly detecting MSCC for three and four  selective retransmissions.  Throughput gain of MSCC method is significant in low and moderate SNR regime.   

\begin{figure}[ht]
\centering \includegraphics[width=9cm]{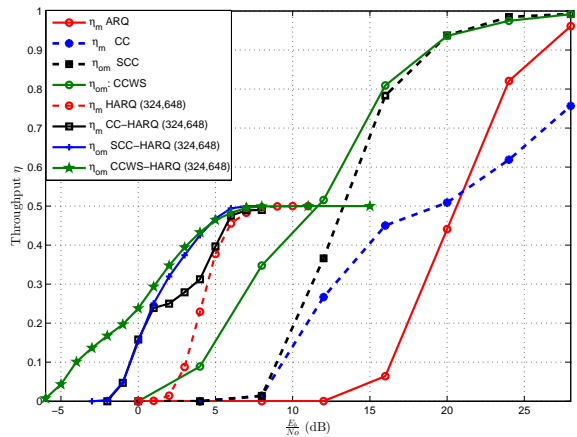}
\caption{Throughput comparison of ARQ, SCC, CCWS and CC Vs HARQ, SCC-HARQ, CCWS-HARQ and CC-HARQ at $\tau=0$, $\tau=\tau_o$ and $\tau=\tau_f$, respectively using half rate LDPC code.}.
\label{fig:EtaSCCwithFEC} 
\end{figure}

Figure~\ref{fig:HBERSCCwithFEC} presents simulated BER performance of conventional retransmission methods and proposed selective retransmission methods with half-rate LDPC (648,324) code.  We denote hybrid SCC and hybrid CCWS  by  SCC-HARQ and CCWS-HARQ, respectively in figure~\ref{fig:HBERSCCwithFEC}. Note that  SCC-HARQ  with threshold  $\tau=0$ is type-I HARQ, which is just single transmission without combining. The threshold  $\tau=\infty$ represents full retransmission and conventional CC-HARQ. In figure~\ref{fig:HBERSCCwithFEC}, $BER_{om}$ of SCC-HARQ represents BER of SCC-HARQ with optimal threshold $\tau_o$ on channel norm that maximizes throughput of the SISO-OFDM system. It is clear from figure~\ref{fig:HBERSCCwithFEC} that in high SNR regime, $BER_{om}$ for SCC-HARQ converges to $BER_{m}$ of single transmission. The joint detection of CCWS-HARQ method with $\tau=\infty$ is equivalent to having four receive antennas. BER of CCWS-HARQ with optimal threshold $\tau_o$ is very close to that of with $\tau=\infty$.

Figure \ref{fig:EtaSCCwithFEC} provides Monte Carlo simulation results of throughput for ARQ, SCC, CCWS,  
HARQ, CC-HARQ, SCC-HARQ and CCWS-HARQ . As it is clear from figure \ref{fig:EtaSCCwithFEC} that in low SNR regime, throughput of conventional and proposed methods with FEC is higher than conventional and proposed methods without FEC.  In high SNR regime, normalized throughput of uncoded methods is higher than coded schemes. Figure \ref{fig:EtaSCCwithFEC} also reveals that throughput of  CCWS has higher throughput as compared to SCC method. Also, proposed methods have higher throughput than conventional coded and uncoded CC methods.   Thus, simulation results demonstrate efficacy of the proposed methods with and without channel coding.

\vspace{-0.3cm}
\section{Conclusion}

\label{Sec:CONCLUSION} We presented performance analysis and throughput
optimization of the proposed selective retransmission methods at modulation layer
for OFDM system. We provide tight BER upper bound  and lower throughput bound for SCC and CCWS methods. We optimize threshold $\tau$ to maximize throughput
of the proposed selective  Chase combining methods. We also generalize SCC method to multiple retransmission at modulation layer in  response to packet failure. The simulation results show
that there is marginal gap between throughput from Monte Carlo runs and that of
from throughput analysis. The simulation results also demonstrate significant
throughput gain of optimized selective retransmission methods over
conventional Chase combining methods with and without channel coding. 
\vspace{0.3cm}

\ifCLASSOPTIONcaptionsoff \newpage{}\fi 
\bibliographystyle{IEEEtran}
\bibliography{report_refrevised}

\end{document}